\documentclass[a4paper]{jpconf}
\usepackage{amssymb}
\usepackage{graphicx}

\newcommand{\be}{\begin{equation}}
\newcommand{\ee}{\end{equation}}

\newcommand{\bea}{\begin{eqnarray}}
\newcommand{\eea}{\end{eqnarray}}

\begin{document}
\title{Box-Cox transformation of firm size data in statistical analysis}

\author{Ting Ting Chen$^{1}$ and Tetsuya Takaishi$^{2}$}
\address{$^{1}$Faculty of Integrated Arts and Sciences, Hiroshima University, Higashi-Hiroshima 739-8521, Japan}
\address{$^{2}$Hiroshima University of Economics, Hiroshima 731-0192, JAPAN}

\ead{$^{1}$d102355@hiroshima-u.ac.jp}
\ead{$^{2}$tt-taka@hue.ac.jp}

\begin{abstract}
Firm size data usually do not show the normality that is often assumed in statistical analysis 
such as regression analysis.
In this study we focus on two firm size data: the number of employees and sale.
Those data deviate considerably from a normal distribution.  
To improve the normality of those data we transform them 
by the Box-Cox transformation with appropriate parameters.
The Box-Cox transformation parameters are determined 
so that the transformed data best show the kurtosis of a normal distribution.
It is found that the two firm size data transformed by the Box-Cox transformation
show strong linearity. This indicates that the number of employees and sale have
the similar property as a firm size indicator.
The Box-Cox parameters obtained for the firm size data are found to be
very close to zero. In this case  the Box-Cox transformations are approximately 
a log-transformation. This suggests that the firm size data we used 
are approximately log-normal distributions. 
\end{abstract}

\vspace{-5mm}
\section{Introduction}
In social science we often use statistical techniques 
such as regression analysis to find relationships among data and 
to make predictions.
In many cases to utilize statistical techniques  normality of 
the data is assumed.
In practice, however, the normality assumption is often violated.
The violation of the normality causes certain difficulties\cite{Sakia,Hossain}.
One way to overcome these difficulties   
is to transform variables to those having desired properties.
In this study we use the Box-Cox power transformations\cite{BoxCox} that also 
include a log-transformation as a special case.

Firm size data are important variables to find relationships 
among financial indicators.
However it is well-known that 
firm size data are not normally distributed and often suggested 
to follow a log-normal distribution\cite{Gibrat,Kalecki,Hart,Quandt,Steindl}.
On the other hand there also exist
studies  that claim firm size distributions
deviate from the log-normal distribution\cite{Cabral,Kaizoji}. 

In this study we use two definitions for firm size:
the number of employees and sale.
As analyzed later those firm size data are not normally distributed.
To improve the normality 
we apply the Box-Cox transformation for those data.
The Box-Cox transformation parameters are determined 
so that the kurtosis of the transformed data comes close 
the kurtosis of  a normal distribution.
The similar approach that uses the skewness has been taken to determine 
the optimum parameter for the Box-Cox transformation\cite{Osborne}.

\section{Box-Cox Transformation}
Let $x_i$ be the i-th observations.
The Box-Cox transformation of $x_i$ is given by 
\be
x(\lambda)_i = \left\{ 
\begin{array}{lll}
\frac{(x_i+c)^\lambda-1}{\lambda}  &  \mbox{if  } & \lambda \neq 0 \\
\log(x_i+c)  &  \mbox{if  } & \lambda = 0 
\end{array}
\right.
\ee
where $c$ is a constant to ensure that $x_i+c$ is positive.
Since firm sizes are all positive we set $c=0$.

\section{Firm Size Data}
We use two firm size data: the number of employees and sale that
consist of 3206 companies traded on the Tokyo Stock Exchange in 2011.
Figs 1 and 2 show the number of employees and sale for 
3206 companies. It is seen  that they are broadly distributed.
Figs 3 and 4 show distributions of the two firm size data.
It is noticed that they are not normally distributed. 

\section{Box-Cox Transformed Data}
In order to find the optimum Box-Cox transformation parameter that best exhibits the kurtosis
of a normal distribution we calculate kurtosis as a function of $\lambda$.
Fig.5 shows the kurtosis as a function of $\lambda$.  
Here the kurtosis $\kappa$ is defined by $\kappa = \frac{ E[(x-\mu)^4]}{Var(x)^2}$
where $\mu=E[x]$.
As seen in Fig.5 
it turned out that the kurtosis is always bigger than 3. 
Thus the optimum parameters of the number of employees and sale can be obtained 
at minimum points. 
Then we obtain the optimum parameters and the kurtosis at the optimum parameters
$\kappa$ as $\lambda_c=0.005$ and $\kappa=3.69$  for the number of employees, and
$\lambda_c=0.022$ and $\kappa=3.77$ for sale.

\begin{figure}[t]
\begin{minipage}{0.5\hsize}
\begin{center}
\vspace{4mm}
\includegraphics[height=5cm]{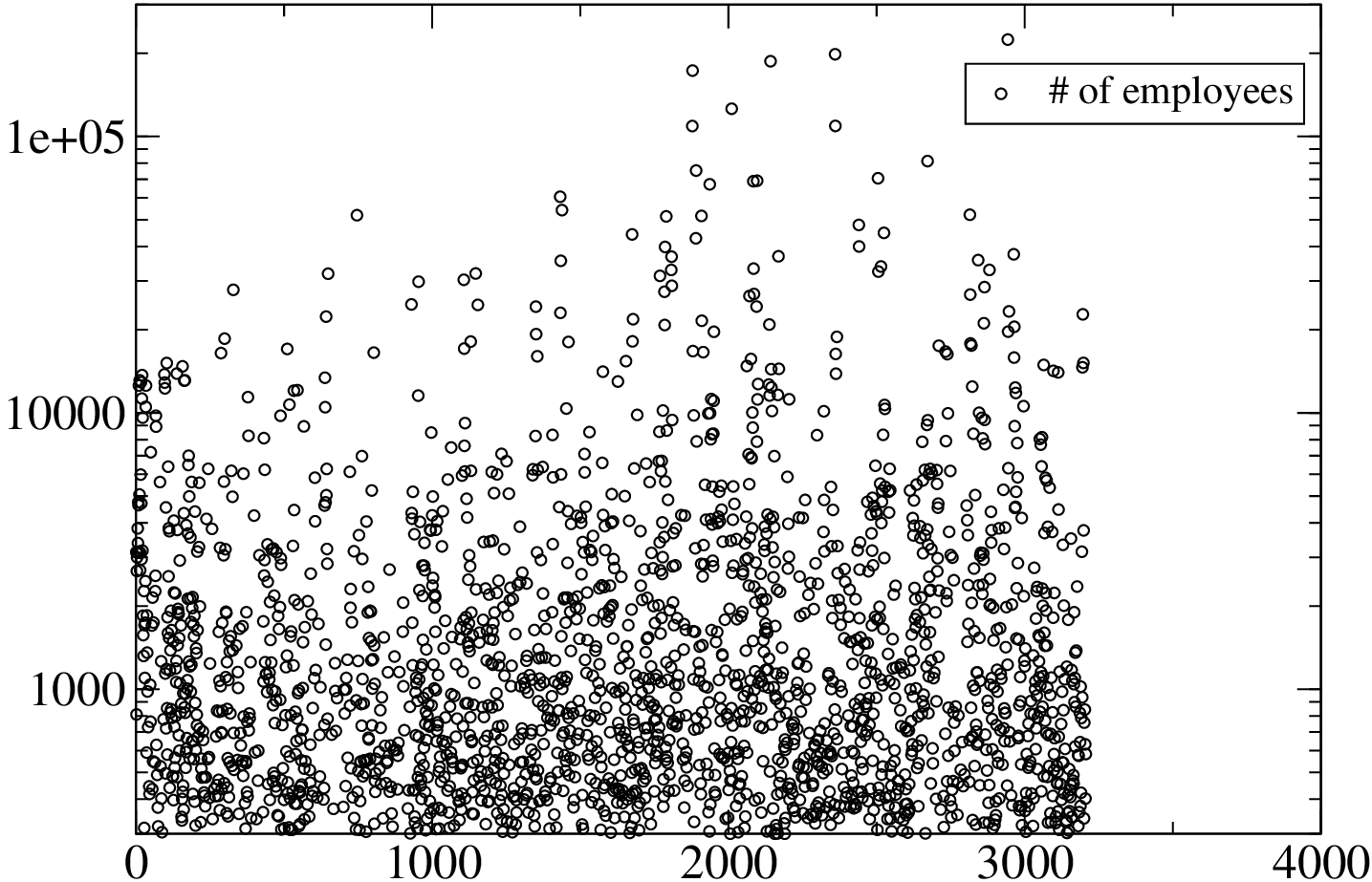}
\caption{ 
The number of employees for 3206 companies.
}
\label{fig:dH}
\end{center}
\end{minipage}
\hspace{3mm}
\begin{minipage}{0.5\hsize}
\begin{center}
\vspace{-1mm}
\includegraphics[height=5cm,keepaspectratio=true]{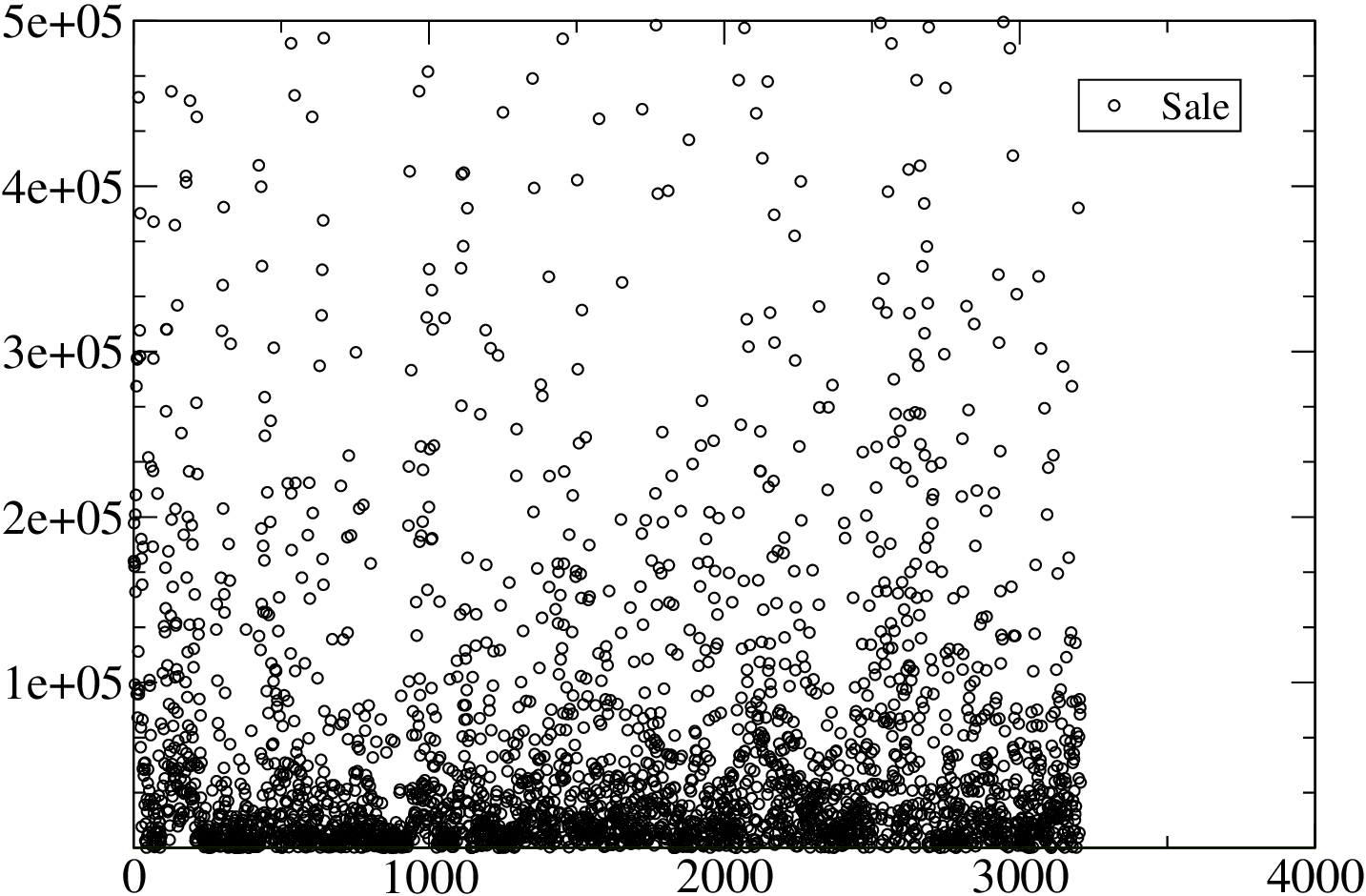}
\caption{The same as in Fig.1 but for sale. \\ \\
}
\label{fig:Acc}
\end{center}
\end{minipage}
\vspace{2mm}
\end{figure}

\begin{figure}[t]
\vspace{5mm}
\begin{minipage}{0.5\hsize}
\begin{center}
\includegraphics[height=5cm]{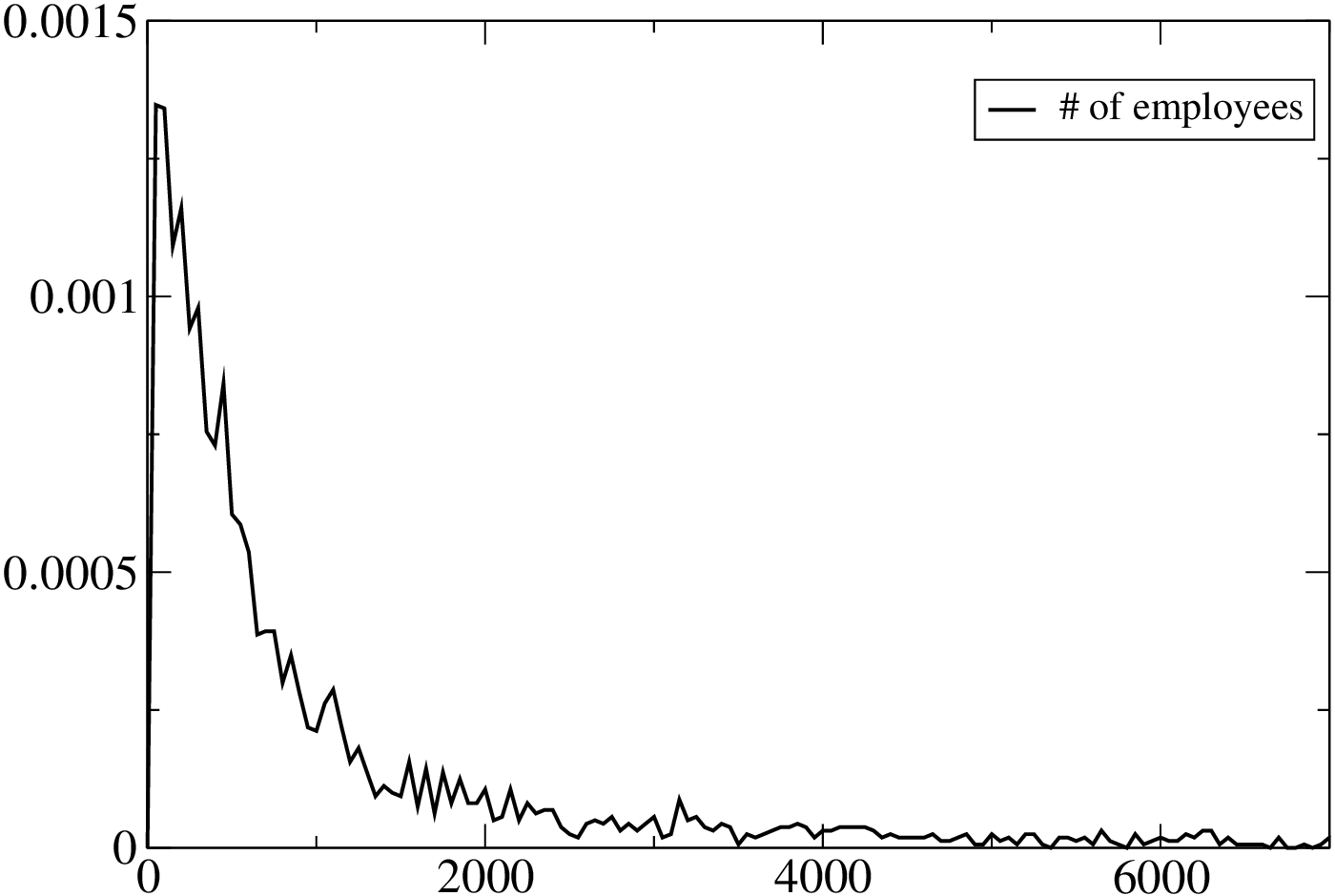}
\caption{ 
Distribution of the number of employees.
}
\label{fig:dH}
\end{center}
\end{minipage}
\hspace{3mm}
\begin{minipage}{0.5\hsize}
\begin{center}
\vspace{-1mm}
\includegraphics[height=5cm,keepaspectratio=true]{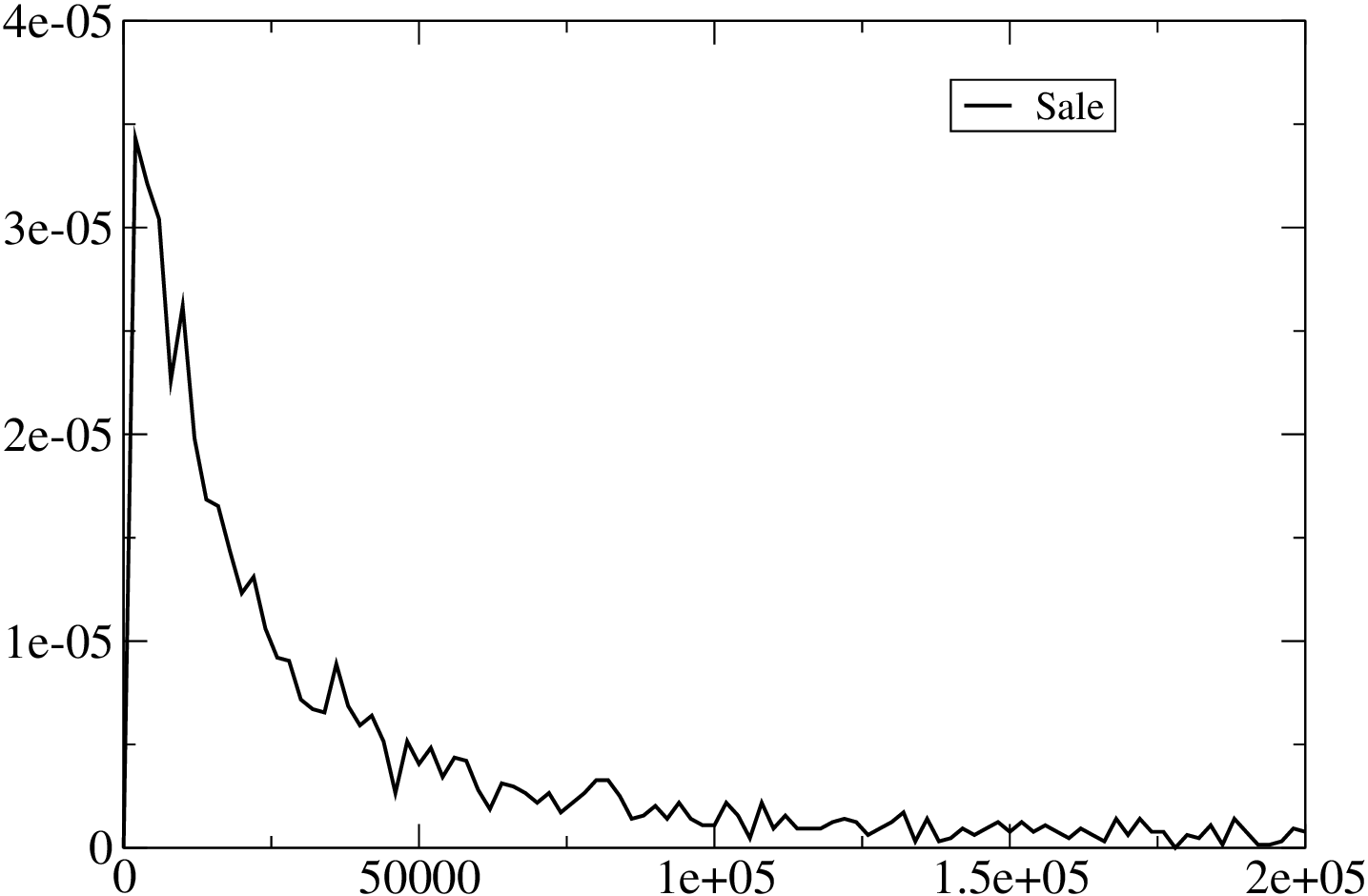}
\caption{
Distribution of sale. \\
}
\label{fig:Acc}
\end{center}
\end{minipage}
\end{figure}

\begin{figure}[t]
\vspace{3mm}
\begin{center}
\includegraphics[height=5cm]{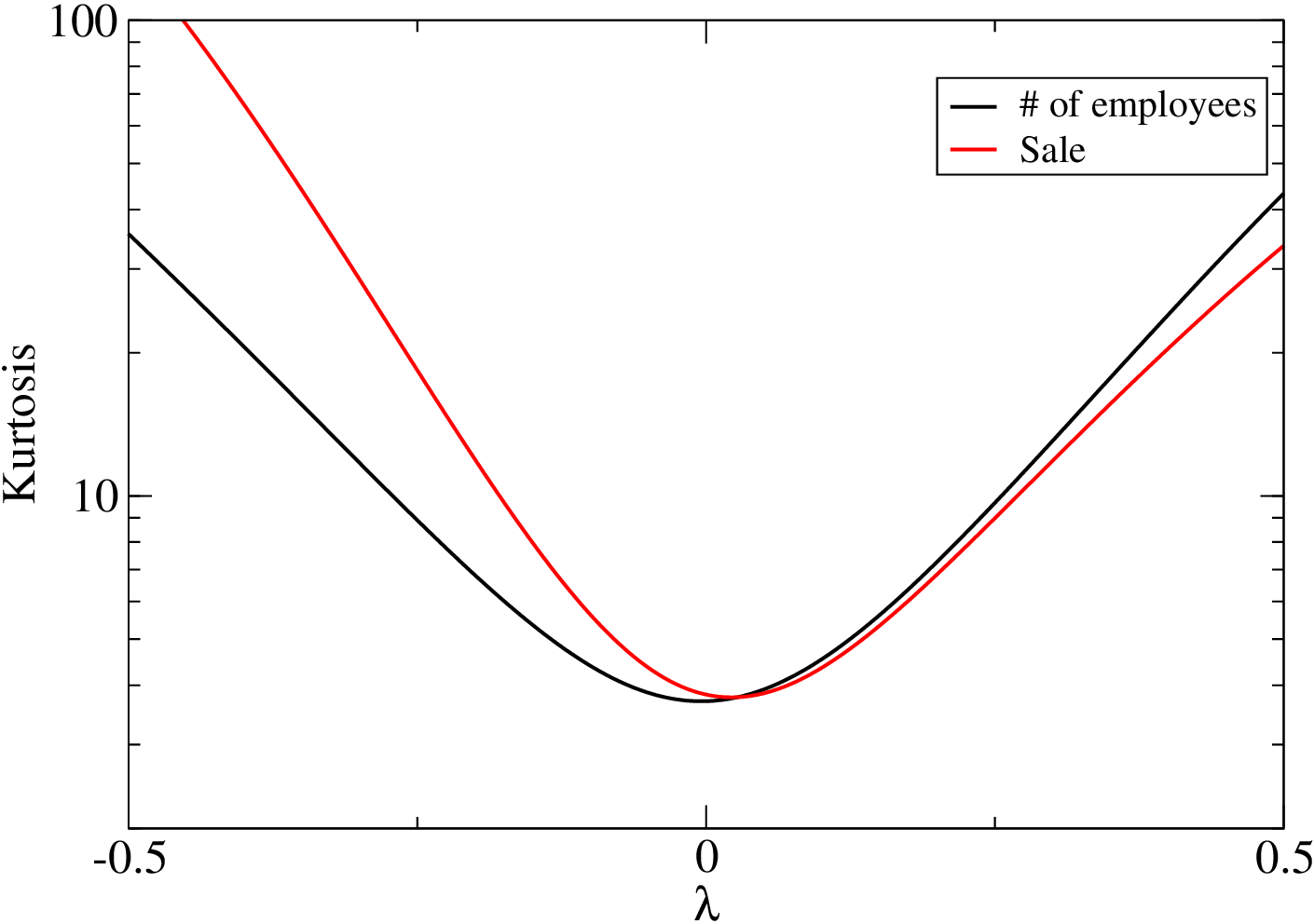}
\caption{ 
Kurtosis as a function the Box-Cox transformation parameter $\lambda$.
}
\end{center}
\end{figure}


Figs 6 and 7 show distributions of the number of employees and sale after the Box-Cox transformation
respectively.
It is clearly seen that the distributions are transformed to more normal ones.  
The red lines in the figures are the fitting results to a normal distribution.


\begin{figure}[t]
\begin{minipage}{0.5\hsize}
\begin{center}
\vspace{4mm}
\includegraphics[height=5cm]{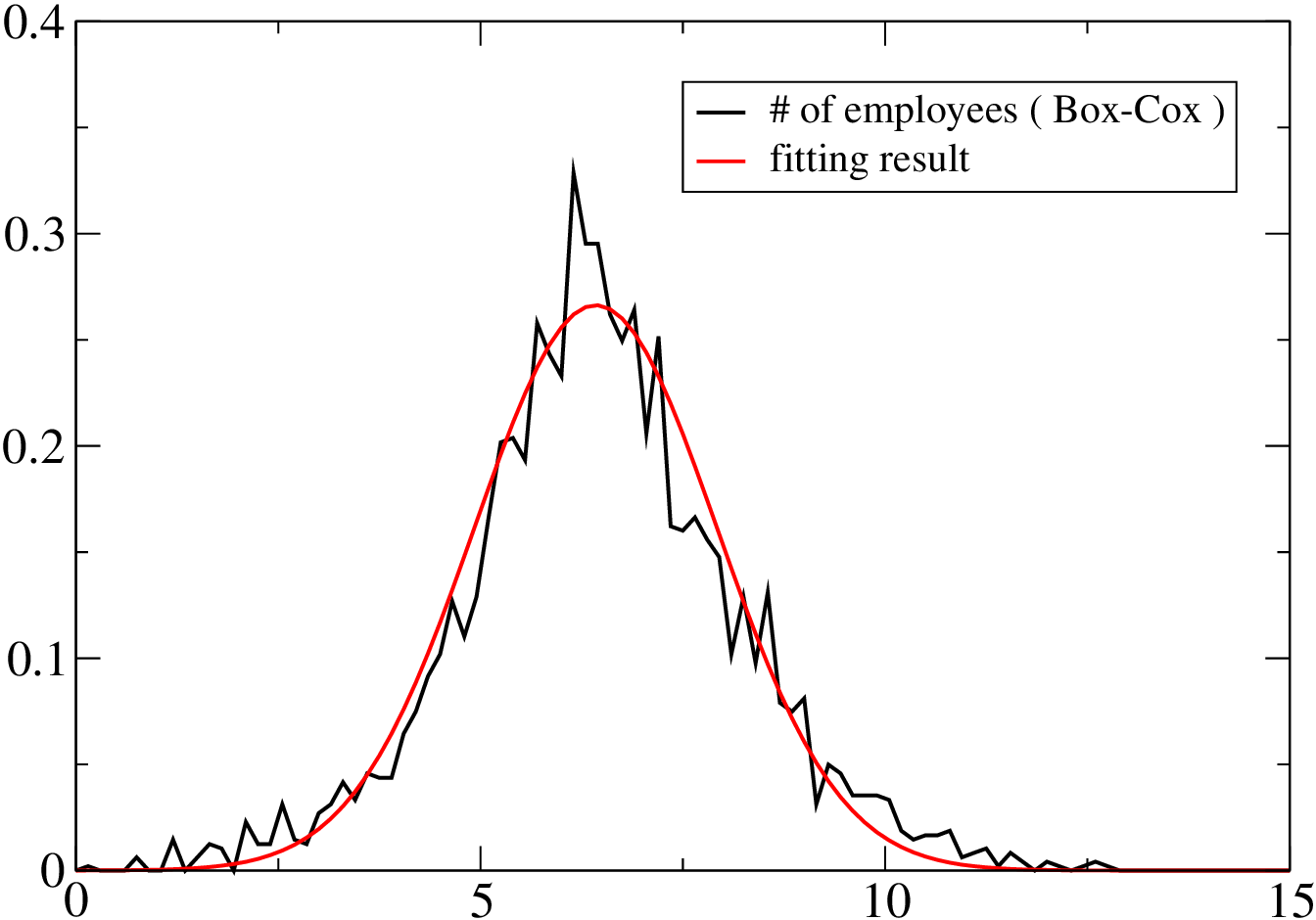}
\caption{Distribution of the number of employees
after the Box-Cox transformation with the optimum parameter. 
The red line is the fitting result to $\exp(-(x-\mu)^2/(2\sigma^2))/(2\pi \sigma^2)^{1/2}$
with $\sigma^2=2.24$ and $\mu=6.42$.
}
\label{fig:dH}
\end{center}
\end{minipage}
\hspace{3mm}
\begin{minipage}{0.5\hsize}
\begin{center}
\vspace{-6mm}
\includegraphics[height=5cm,keepaspectratio=true]{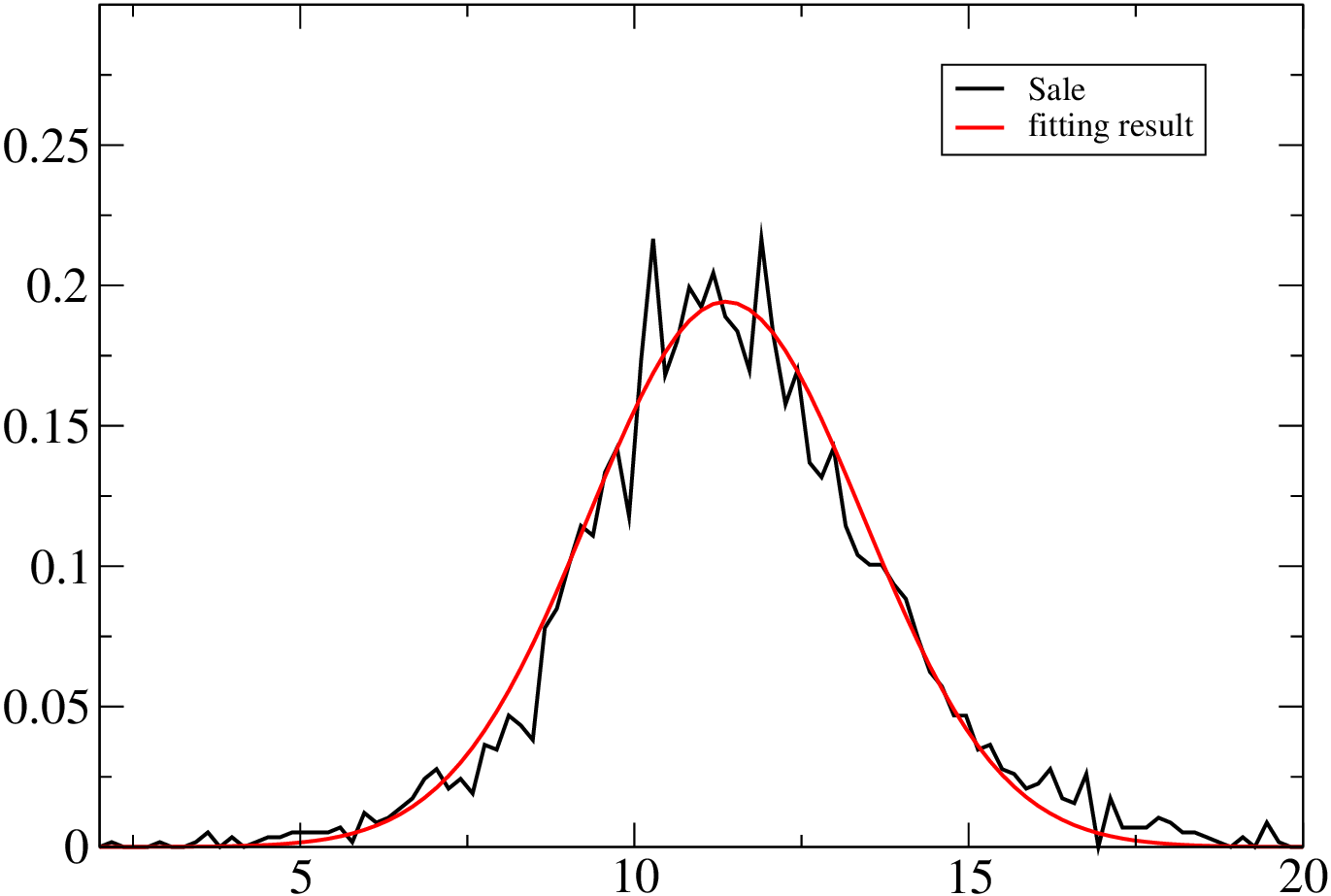}
\caption{
The same as in Fig.8 but for sale with $\sigma^2=4.22$ and $\mu=11.37$.
}
\label{fig:Acc}
\end{center}
\end{minipage}
\end{figure}

Next we investigate a relationship between 
the number of employees and sale.
There is no obvious reason that they are closely related.
In fact as seen in Fig.8 no clear relation is seen 
between the number of employees and sale. The correlation coefficient between them
is calculated to be 0.662.
On the other hand Fig.9 shows the relation between the number of employees and sale
after the Box-Cox transformation which indicates that 
the number of employees and sale have a strong linearity. 
The  correlation coefficient is calculated to be 0.873 that also supports the strong linearity.
This result may suggest that the number of employees and sale can be used 
as a firm size indicator that has the similar property.

\begin{figure}[t]
\begin{minipage}{0.5\hsize}
\begin{center}
\includegraphics[height=5cm]{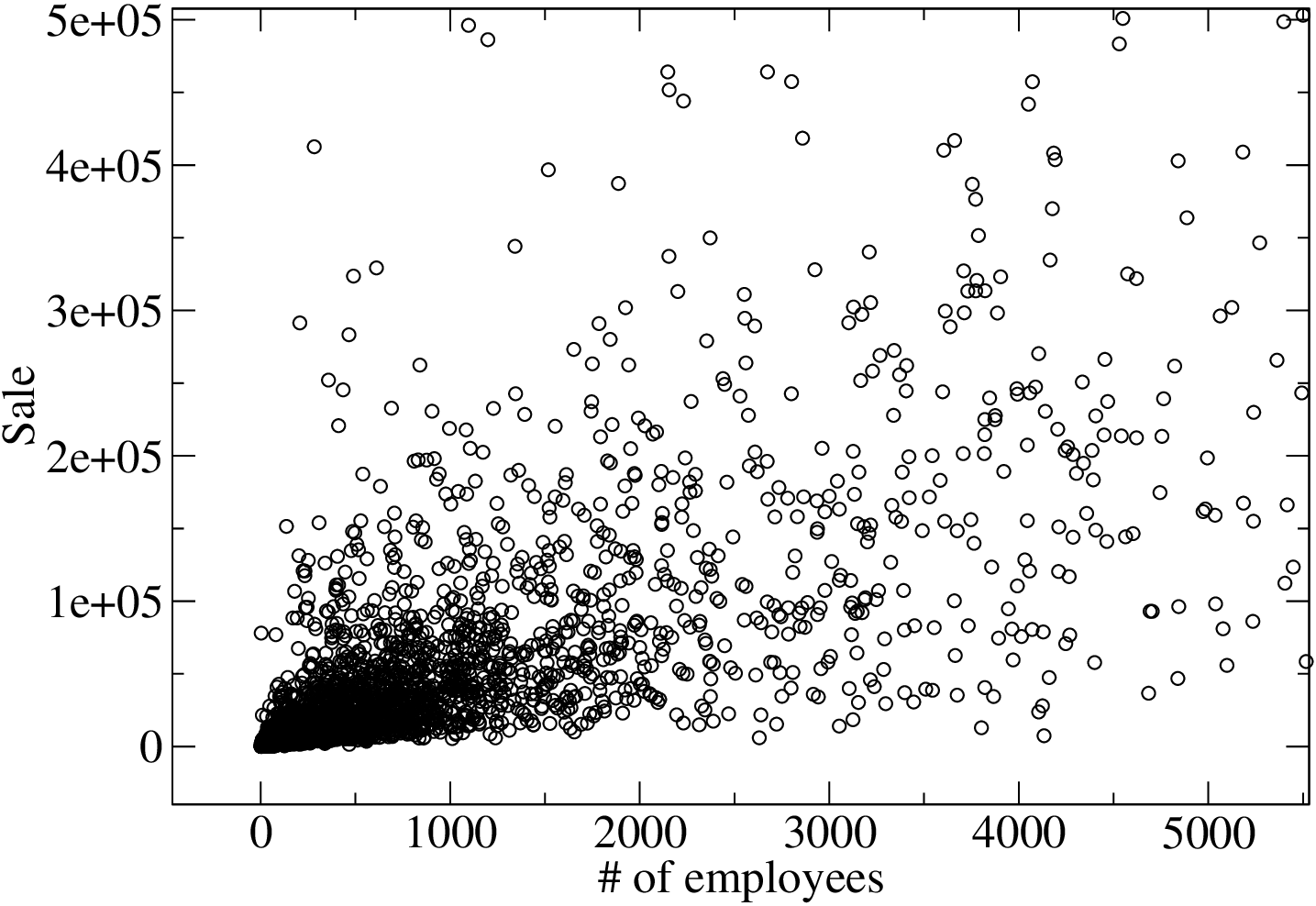}
\caption{The number of employees versus sale.
}
\label{fig:dH}
\end{center}
\end{minipage}
\hspace{3mm}
\begin{minipage}{0.5\hsize}
\begin{center}
\vspace{8mm}
\includegraphics[height=5cm,keepaspectratio=true]{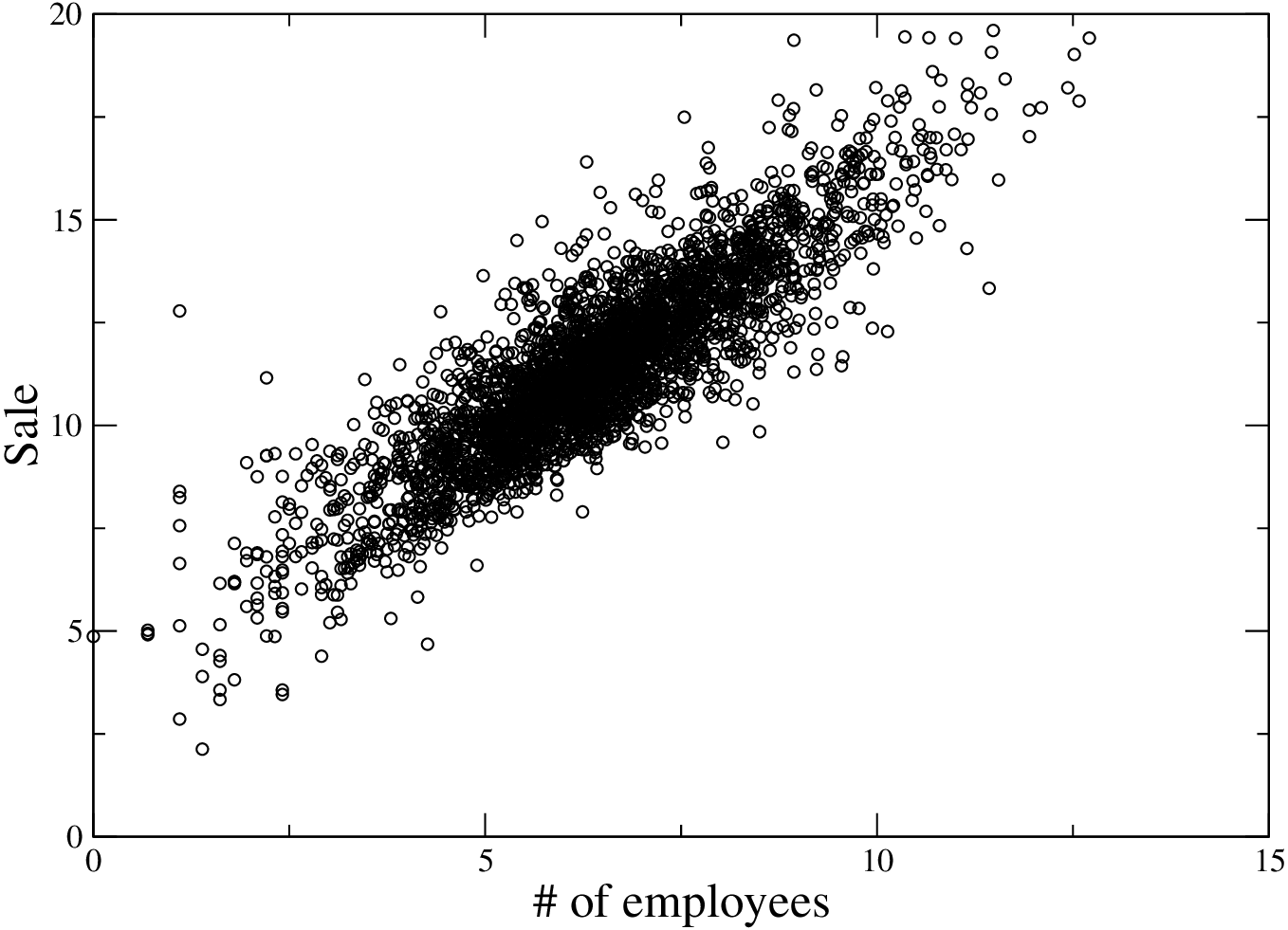}
\caption{
The number of employees versus sale,
after the Box-Cox transformation with the optimum parameter.
}
\label{fig:Acc}
\end{center}
\end{minipage}
\end{figure}

\section{Conclusion}
In order to improve the normality of firm size data
we have used the Box-Cox transformation.
The optimum Box-Cox transformation parameters are 
determined so that the transformation brings the data 
to more normal ones.
We find that the number of employees and sale after the Box-Cox transformation 
show the strong linear relation.
This may indicate that the number of employees and sale has the similar property 
as a firm size indicator.

The optimum Box-Cox transformation parameters we obtained are very close to zero.
Thus the Box-Cox transformations we performed are actually close to a log-transformation,
which means that distributions of the number of employees and sale are close 
to a log-normal distribution. 
This is consistent with the previous literature which claims that 
the firm size distributions follow the log-normal distributions\cite{Gibrat,Kalecki,Hart,Quandt,Steindl}.

\section*{Acknowledgement}
Numerical calculations in this work were carried out at the
Yukawa Institute Computer Facility.
This work was supported by JSPS KAKENHI Grant Number 25330047.


\section*{References}
\begin{thebibliography}{9}


\bibitem{Sakia}
Sakia R M  1992
\textit{The Statistician} {\bf 41} 169-178

\bibitem{Hossain}
Hossain M Z 2011
\textit{Journal of Emerging Trends in  Economics and Management Sciences} {\bf 2(1)} 32-39

\bibitem{BoxCox}
Box G E P and Cox D R 1964
\textit{Journal of the Royal Statistical Society B} {\bf 26} 211-234

\bibitem{Gibrat}
Gibrat R 1931
\textit{Les Inegalites Economiques}


\bibitem{Kalecki}
Kalecki M 1945
{\em Econometrica} {\bf 13} 161-170

\bibitem{Hart}
Hart P E and Paris S J 1956
{\em Economica (New Series)} {\bf 29} 29-39


\bibitem{Quandt}
Quandt R E 1966
{\em American Economic Review} {\bf 56} 416-432

\bibitem{Steindl}
Steindl J 1965
{\em Random Process and the Growth of Firms} New York, Hafner



\bibitem{Cabral}
Cabral L M B and Mata J 2003
{\em  American Economic Review} {\bf 93} 1075-1090

\bibitem{Kaizoji}
Kaizoji T, Iyetomi H and Ikeda Y 2006
\textit{Evolutionary and Institutional Economics Review} {\bf 2(2)} 183-198

\bibitem{Osborne}
Osborne J W 2010 
\textit{Practical Assesment, Research \& Evaluation} {\bf 15}  No.12 







\end{thebibliography}
\end{document}